\documentclass[
aps,
showpacs,
preprint,
tightenlines,
nofootinbib,
floatfix]
{revtex4}
\usepackage{epsfig}
\usepackage{graphicx,
longtable
}

\begin{document}
\title{Probing neutrino magnetic moment and unparticle interactions with
Borexino}

\author{
Daniele Montanino$^1$, Marco Picariello$^1$, Jo\~ao Pulido$^2$}
\affiliation{
$^1$ Dipartimento di Fisica, Universit\`a del Salento 
and Sezione INFN di Lecce\\
Via Arnesano, I--73100 Lecce, Italy\\
$^2$ Centro de Fisica Teorica das Particulas, Departamento de Fisica\\
Instituto Superior Tecnico 1049--001 Lisboa, Portugal\\}

\begin{abstract} 
We discuss the limits on the neutrino magnetic moment and hypothetical
interactions with a hidden unparticle sector, coming from the first
neutrino data release of the Borexino experiment. The observed spectrum
in Borexino depends weakly on the solar model used in the analysis,
since most of the signal comes from the mono-energetic $^7$Be neutrinos.
This fact allows us to calibrate the $\nu$-$e$ scattering cross section
through the spectral shape. In this way, we have derived a limit on the
magnetic moment for the neutrinos coming from the Sun (in which a
$\nu_\mu$ and $\nu_\tau$ component is present): $\mu_\nu \leq
8.4\times10^{-11}\mu_B$ (90\%CL) which is comparable with those obtained
from low energy reactor experiments. Moreover, we improve the previous
upper limit on magnetic moment of the $\nu_\tau$ by three orders of
magnitude and the limit on the coupling constant of the neutrino with a
hidden unparticle sector. 
\end{abstract}


\maketitle

\section{Introduction}

The present experimental world neutrino data (apart the LSND experiment
which was not confirmed by the recent MiniBooNE result) provide a robust
interpretation in terms of a three active neutrino oscillation scenario
(for a recent review, see e.g.\ \cite{Schwetz:2007my}). In particular,
solar neutrinos together with KamLAND data, can be explained in a
simplified two--neutrino framework (in the limit of a vanishing
$\theta_{13}$) with \cite{Shirai:2007zz}:
\begin{eqnarray}\label{eq:massmixing}
\Delta m_{12}^2 = \left(7.58\pm^{0.21}_{0.20}\right)\times 10^{-5}\, 
{\rm eV}^2\, ,\nonumber\\ 
\tan^2 \theta_{12} = 0.56\pm^{0.14}_{0.09}\, .
\end{eqnarray}
However, there is still room for subleading non standard neutrino
interactions in the interpretation of data. For example, an evidence for
time modulation of the oscillation probability in the solar neutrinos
could be explained in terms of a non-zero magnetic moment which could
induce transitions among active and sterile neutrinos
\cite{Picariello:2007sw}.

Recently, the Borexino collaboration has released the first data
relative to about three months of data taking
\cite{Collaboration:2007xf}. In this experiment, solar neutrinos
(mainly, those coming from the $^7$Be source) are detected through
$\nu$-$e$ scattering and the recoil electron energy is measured through
a scintillation technique. The observed event rate is essentially
consistent with the one predicted by the Standard Solar Model
\cite{Bahcall:2005va} and the oscillation hypothesis. However, since the
main source observed in Borexino is the monoenergetic 863~keV $^7$Be
source, a precise calibration of the differential
$d\sigma/dT_e(E_\nu,T_e)$ cross section (where $E_\nu$ is the incident
neutrino energy and $T_e$ the recoil electron kinetic energy) is
possible through a spectral shape analysis. For example, neutrino
electromagnetic form factors would influence the scattering cross
section \cite{Vogel:1989iv}. In particular, a non zero neutrino magnetic
moment introduces a term which grows with the inverse of both the energy
of the incident neutrino and with that of the recoil electron. For this
reason, a low energy experiment (such as Borexino), is in a favorable
situation.

Although the limits obtained are still weaker than those obtained by a
direct measurement of the $\bar{\nu}_e$-$e$ scattering in reactor
experiments \cite{Wong:2006nx,Beda:2007hf} (and those that could be
obtained by Borexino itself from a calibration experiment with an
external source of neutrinos or antineutrinos \cite{Ianni:1999nk}) we
should stress that these are short baseline experiments, and they
measure the magnetic moment of the $\bar{\nu}_e$ component. Instead,
solar neutrinos embed also a component of $\nu_\mu$ and $\nu_\tau$, for
which the limits are much weaker \cite{PDG,Raidal:2008jk}. 
SuperKamiokande data were used in the past to study the neutrino magnetic 
moment
\cite{Mourao:1992ip,Pulido:1997tx,Beacom:1999wx,Grimus:2002vb,Liu:2004ny}, 
but since
this experiment observes the continuous $^8$B source, it is difficult to
disentangle the effects of a spectral distortion due to non--standard
interactions and those due to the oscillation mechanism. In the past,
the Borexino collaboration tried to put a limit on $\mu_\nu$ using the
prototype of the Borexino detector (the Counting Test Facility, CTF)
\cite{Back:2002cd}. However at that time, due to the smallness of the
CTF, no solar neutrinos were observed and a limit was established
assuming the theoretical SSM neutrino flux.

Concerning other non--standard interactions, the possibility of a
conformal hidden sector, called ``unparticle sector'', which couples to
the various gauge and matter fields of the SM through non-renormalizable
interactions has been recently proposed \cite{Georgi:2007ek}. The
unparticle sector is assumed to have a non-trivial infrared fixed point,
$\Lambda_{\cal U}$, below which the sector has a scale invariance and
the hidden operators become an effective unparticle operator with
non-integral scaling dimension $d$. Limits from low-energy
neutrino-electron scattering in the unparticle physics framework have
been recently obtained in \cite{Balantekin:2007eg}. We will show that
the limits coming from Borexino are stronger.

The plan of the paper is the following. In Sec.~\ref{sec:interactions}
we review the contribution to the cross section due to the neutrino
magnetic moment or the coupling with the unparticle sector; in
Sec.~\ref{sec:Expinput} we describe briefly the experimental input; in
Sec.~\ref{sec:analysis} we describe our analysis technique; in
Sec~\ref{sec:results} we derive the upper bounds on the neutrino
magnetic moment from SK and Borexino experiments, emphasizing the fact
that the former is quite sensitive to the Solar Model assumed while the
latter is independent; finally, in Sec.~\ref{sec:conclusions} we draw
our conclusions.

\section{Electron-neutrino cross section}\label{sec:interactions}

\subsection{Neutrino magnetic moment}

For a neutrino with flavor $a$, the scattering process standard
differential $\nu_a$-$e$ cross section as a function of the incident
neutrino energy $E_\nu$ and of the recoil electron kinetic energy $T_e$,
is given by \cite{tHooft}:%
\footnote{In this work for completeness we have also included the 1-loop
corrections to this formula \cite{Bahcall:1995mm}. However, the effect
of such corrections is negligible.}
\begin{equation}\label{eq:weakcross}
\frac{d\sigma^{\rm std}_a}{dT_e}(E_\nu,T_e)=
\frac{\sigma_0}{m_e}\left[(g^a_V+g^a_A)^2+
(g^a_V-g^a_A)^2\left(1-\frac{T_e}{E_\nu}\right)^2- \left((g^a_V)^2 -
(g^a_A)^2\right)\frac{m_eT_e}{E_\nu^2}\right]\, ,
\end{equation}
with $\sigma_0=G_F^2 m^2_e/(2\pi)$. For $\mu$ and $\tau$ neutrinos,
where only neutral current interactions are possible, the Standard Model
of electroweak interactions provides $g^{\mu,\tau}_V=
2\sin^2\theta_W-\frac{1}{2}$ and $g^{\mu,\tau}_A=-1/2$, with
$\sin^2\theta_W=0.23122$ \cite{PDG}. For electron neutrinos, where also
charge current interactions are possible, we have $g^e_{V,A}\to
g^{\mu,\tau}_{V,A}+1$.

Besides standard interactions, neutrinos can couple with photons through
a possible magnetic dipole and/or charge radius. The effective low-energy 
$\nu$-$\gamma$ interaction vertex is \cite{Vogel:1989iv}:
\begin{equation}\label{eq:emlagr}
(\Gamma^\rho_{\nu-\gamma})_{ab}=\bar\nu_b\left[
\frac{\langle r_\nu^2\rangle_{ab}}{6} q^2\gamma^\rho
-\frac{1}{2 m_e}\left(\mu_{ab}+d_{ab}\gamma^5\right)
\sigma^{\rho\lambda}q_\lambda\right]\nu_a
\, ,\end{equation}
where $\langle r_\nu^2\rangle_{ab}$ is the charge radius and $\mu_{ab}$ 
and $d_{ab}$ are the neutrino magnetic and electric dipole moments 
respectively. Since for ultrarelativistic neutrinos it is not possible 
to distinguish between the two dipole moments, for simplicity we consider 
only magnetic moments. For an individual incoming neutrino with flavor $a$,
since the outcoming neutrino flavour is in general not observable, the only phenomenologically relevant parameter is a combination of the magnetic moment 
matrix $\mu_{ab}$:
\begin{equation}\label{eq:mu}
\mu_a=\sqrt{\sum_b^{} |\mu_{ab}|^2}\, .
\end{equation}
For Dirac neutrinos $\mu_{ab}$ is a generic complex matrix and involves 
transitions among left and right (sterile) states. Conversely, for Majorana 
neutrinos the transitions are among neutrino and antineutrino states of 
different flavours. In this case  the matrix $\mu_{ab}$ is antisymmetric 
(and in particular, $\mu_{aa}=0$).

Actually, the charge radius could be absorbed in a redefinition 
of the $g_V$'s:
\begin{equation}\label{eq:gvshift}
g_V^a \to g_V^a+\frac{2M_W^2}{3}\langle r_\nu^2\rangle_a\sin^2\theta_W
\, .\end{equation}
However, as we will comment later, Borexino is largely insensitive to
variations on the axial and vector couplings $g^a_V$ and $g^a_A$ within
the limits coming from present phenomenology \cite{Barranco:2007ej}. For
this reason, we limit our analysis to the magnetic moment assuming
standard values for $g^a_V$ and $g^a_A$.

Since the neutrino flavor composition in the experiments considered here
depends on the energy range of each experiment, the bounds we derive in
this work are actually applicable to an effective neutrino magnetic
moment, which is a linear combination of the individual flavor magnetic
moments whose coefficients depends, as will be seen later, on the
weighted average survival probability for each experiment.

Robust cosmological arguments show that $\mu_\nu$ should be smaller than
$10^{-8}$ Bohr magnetons, $\mu_B$, \cite{Mirizzi:2007jd}, although other
astrophysical arguments largely override this bound (see, e.g.,
\cite{Raffelt:1998xu} and reference therein). However, such arguments
are model dependent and thus less reliable. The strongest direct bounds
on the $\bar{\nu}_e$ come from the TEXONO experiment \cite{Wong:2006nx},
i.e., $\mu_e<0.74\times 10^{-10}\mu_B$ and, more recently, from the
GEMMA experiment, i.e., $\mu_e<0.58\times 10^{-10}\mu_B$ at $90\%$ C.L.
\cite{Beda:2007hf}. However the limits for the $\mu_{\mu,\tau}$ are much
weaker ($\mu_\mu<6.8\times 10^{-10}\mu_B$ \cite{Auerbach:2001wg},
$\mu_\tau<3900\times 10^{-10}\mu_B$ \cite{Schwienhorst:2001sj}).

The contribution to the $\nu_a$-$e$ cross section due to the neutrino
magnetic moment interaction is given by \cite{Vogel:1989iv}:
\begin{equation}\label{eq:mucontib}
\mu_a^2\frac{d\sigma^\mu}{dT_e}(E_\nu,T_e)=\frac{\pi \alpha_{\rm
e.m.}^2}{m_e^2} \left(\frac{\mu_a}{\mu_B}\right)^2
\left(\frac{1}{T_e}-\frac{1}{E_\nu}\right)\, ,
\end{equation}
where we have explicitly factorized out the $\mu_a$ dependence from the
expression of the cross section. A comment is in order. In principle an
interference term between the magnetic moment and the weak interaction
is possible. However, this interference term vanishes if the neutrino
are longitudinally polarized and the electrons are unpolarized. If
neutrinos cross a strong magnetic field (such as the solar one) they can
acquire a transverse polarization due to the precession. In this case an
interference effect could contribute to the cross section
\cite{Vogel:1989iv}. Here we neglect this effect since the first
Borexino data was taken in a period of low magnetic field activity.

\subsection{Coupling with Unparticles}

Recently, a scale invariant (``unparticle'') sector which decouples at
high energy was proposed in \cite{Georgi:2007ek}. Leptons can couple for
example with a scalar unparticle sector
\cite{Georgi:2007ek,Cheung:2007ue,Zhou:2007zq} through the Lagrangian%
\footnote{In general it is possible also the case in which a coupling
with a vector or tensor unparticle is considered. Here, for simplicity, we
consider only the scalar case.} 
\begin{equation}
{\cal L}_{\cal U}=\lambda_e\frac{1}{\Lambda_{\cal U}^{d-1}}\bar{e}
\hat{O}_{\cal U} e + 
\sum_{a,b}\lambda_\nu^{a b}\frac{1}{\Lambda_{\cal U}^{d-1}} 
\bar{\nu_a} \hat{O}_{\cal U}\nu_b+ {\rm h.c.}\, , 
\end{equation}
where $d$ is the non-integral scaling mass dimension of the unparticle
operator, $\Lambda_{\cal U}$ is a typical scale of the unparticles
physics and it can be assumed $\sim O($TeV$)$, $\hat{O}_{\cal U}$ is the
unparticle operator, and the $\lambda$'s are the coupling constants of
the leptons to the unparticle sector (possible flavor changing
interactions $\nu_a\to\nu_b$ have been also taken into account). The
contribution to the scattering amplitude for elastic $\bar \nu_a$-$e$
scattering from the exchange with a scalar unparticle is
\cite{Balantekin:2007eg}
\begin{equation}\label{eq:unpscat}
{\cal M}^{a b}=\lambda_\nu^{a b}\lambda_e\frac{{\cal
F}(d)}{\Lambda_{\cal U}^{2d-2}} \left[\bar{\nu}_b(k_f) \nu_a(k_i)\right]
\frac{1}{(-q^2)^{2-d}}\left[\bar{e}(p_f) e(p_i)\right]\, , 
\end{equation}
where $q=k_f-k_i$ and
\begin{equation}
{\cal F}(d)=\frac{8\pi^{5/2}}{{(2\pi)}^{2d}\sin (\pi d)}
\frac{\Gamma(d+1/2)}{\Gamma(d-1)\Gamma(2d)}\,.
\end{equation}
From this amplitude, the contribution to the $\nu_a$-$e$ cross section
can be calculated:%
\footnote{One can easily check that for almost massless neutrinos the
interference term with the standard amplitude is negligible.}
\begin{equation}\label{eq:unpartcs}
\lambda_a^2\frac{d\sigma^{\cal U}}{dT_e}(E_\nu,T_e)=
\lambda_a^2\frac{2^{2d-7}{\cal F}^2(d)}{\pi E_\nu^2 \Lambda_{\cal
U}^{4d-4}} (m_e T_e)^{2d-3}(T_e+2m_e)\, ,
\end{equation}
and we have conveniently defined
\begin{equation}
\lambda_a=\sqrt{\sum_b^{} (|\lambda_\nu^{a b}|\lambda_e)^2}\,.
\end{equation}
As in the case of magnetic moment, the final state will be a sterile
right state or an antineutrino state depending on the Dirac or Majorana
nature of the neutrinos. Notice that for $d<3/2$ the cross section diverges 
for low $T_e$, thus low energy experiments are most sensitive to the 
unparticles. Notice that for $d=1$ we have the same $T_e^{-1}$ dependence 
as in the case of the magnetic moment.

\section{The experimental input}\label{sec:Expinput}

The Borexino experiment at the Gran Sasso underground laboratory is
designed to study mainly the 863~keV monoenergetic $^7$Be solar
neutrinos, through a real-time and low-background detector. A detailed
description of the experimental apparatus and its ancillary plants can
be found in \cite{Borexinodetector}. Briefly, Borexino consists of
300~tons of an high purity liquid organic scintillator (pseudocumene,
C$_{9}$H$_{12}$, $\rho$=0.88~g/cm$^3$) doped with PPO at 1.5~g/l. The
scintillator mixture is contained inside a thin nylon sphere (8.5~m in
diameter). This volume is viewed by 2212 8'' photomultipliers (PMTs)
installed on a stainless steel sphere of 13.7~m in diameter. The
scintillator is shielded against background from PMTs and other external
sources by a 2.5~m buffer of pseudocumene, an outer nylon vessel and 2~m
of high purity water. For solar neutrinos search a fiducial volume of
100~tons is selected off-line. In Borexino solar neutrinos are detected
through the scattering reaction $\nu+e\to\nu+e$. The recoil electron
energy is converted into light inside the scintillator. The intrinsic
$^{14}$C contamination%
\footnote{The $^{14}$C have a $\beta$ decay with an end-point energy of
156~keV.} 
and finite energy resolution set the detection threshold at about
200~keV. In Borexino no directionality is possible to search for
neutrino interactions and it is not possible to distinguish on an
event-by-event basis between neutrino processes and $\beta$/$\gamma$
backgrounds. Therefore, the radiopurity of the scintillator is a
fundamental experimental issue. The first results
\cite{Collaboration:2007xf} have shown that the radiopurity achieved is
beyond the expectations and this allows to extend the research program
as we attempt to do in this work.

The data taking started in May 2007. The collaboration released the
first data in August 2007 \cite{Collaboration:2007xf}. The observed flux
of $^7$Be neutrinos is (within the errors) consistent with the Standard
Solar Model prediction in the hypothesis of
Mikheyeev-Smirnov-Wolfenstain oscillations, i.e., $47\pm
7$(stat.)$\pm12$(sys.) counts per day (cpd) in 100~tons for the $^7$Be
(863~keV) neutrinos, against a theoretical value of 49$\pm$4 cpd's. The
collaboration has also produced a spectrum of the observed events with the
visible electron energy $K_e$ in the range $270\leq K_e\leq 800$~keV. This spectrum
is shown in Fig.~6 of \cite{Collaboration:2007xf}, in which the number
of events per day and per 100~tons of scintillator are shown in 53 bins
of energy (we show our equivalent plot in Fig.~\ref{fig:spectrum}). The
collaboration quote a 15\% uncorrelated error for each bin. However,
since there are also bins with less of three events, the mere
statistical error for these bins would be greater than 15\%. For this
reason we prefer to be conservative and sum in quadrature the 15\% error
quoted by the collaboration to the statistical error $N_i^{-1/2}$, where
$N_i$ is the number of events in each bin. In the majority of bins,
where $N_i\gg 3$, the statistical error is negligible and the
uncorrelated error is just that quoted by the collaboration. The main
source of correlated error comes from the determination of the fiducial
mass. The collaboration quote a 25\% error equally correlated among all
bins.

From \cite{Collaboration:2007xf} we know that the measured light yield
is about 500 photoelectrons/MeV. The light yield affects the energy
resolution of the detector and at a first approximation we can assume a
gaussian energy smearing with $\sigma = \sqrt{T_e/500}$.  We notice that
this assumption does not work well in the low energy regime (mainly
below 200~keV) where non linear effects due to quenching take place. In
the energy range we are considering, namely [270,800]~keV this effect is
expected to be on the order of a few \%'s.  With the above assumption
the resolution function for the detection of solar neutrinos is thus:
\begin{equation}\label{eq:resol}
{\cal R}(K_e, T_e)=\frac{1}{\sqrt{2\pi}\sigma} 
\exp\left[-\frac{\left(K_e-T_e\right)^2}{2\sigma^2}\right]\, ,
\end{equation}
where $T_e$ is the {\em real} kinetic electron energy, $K_e$ is the
visible (measured) one, and $\sigma=4.47\%\sqrt{T_e}$.

The main sources of background in the detector come from the decay of
contaminants contained in the scintillator (being the cosmogenic
contribution almost completely rejected by muon vetoing and other
techniques with the exception of $^{11}$C which, however, could dominate
the spectrum above 1~MeV). Internal background is due to the $\beta$
decays of $^{210}$Bi and $^{85}$Kr and the $\alpha$ decay of $^{210}$Po
as reported in  \cite{Collaboration:2007xf} . The former background  is
enormously reduced by $\alpha/\beta$ pulse shape discrimination (PSD).
However, the rejection efficiency of the PSD is not 100\% and may change
with the energy. In our analysis we adopt the same assumption used in
\cite{Collaboration:2007xf}  and take into account  a possible small
residual of $\alpha$-like events in the $\beta$-like/neutrino events
spectrum by means of a gaussian peaked around 410 KeV:
\begin{equation}
\tilde{S}_\alpha(K_e)={\cal N}_{^{210}{\rm Po}} \cdot {\cal R}(K_e, 410
{\rm KeV})\, .
\end{equation}
Here ${\cal N}_{^{210}{\rm Po}}$ is the unknown normalization to the
spectrum that should be determined by the fit.

As pointed out above $\beta$ decays cannot be rejected. They contribute
to the total spectrum through the 
\begin{equation}
S_\beta(T_e)=\sum_B {\cal N}_B \left(1-\frac{T_e}{Q_B}\right)^2
\frac{E_e p_e}{Q_B^2} \ F_B(T_e)\, , 
\end{equation}
where $B\in \{ ^{210}{\rm Bi},^{85}{\rm Kr}\}$, $E_e=T_e+m_e$ and
$p_e=\sqrt{2 m_e T_e + T_e^2}$ are the total electron energy and
momentum, and the Fermi correction $F_B$ is given by
\cite{Fermicorrected}:
\begin{equation}
F_B(T_e)=2(1 +\gamma_0)\left(2 R p_e\right)^{2\gamma_0-2} e^{\pi
\nu}\frac{|\Gamma(\gamma_0+i\nu)|^2}{\Gamma(2\gamma_0+1)^2}\, ,
\end{equation}
where $R=0.426 A^{1/3}\alpha_{\rm e.m.}/m_e$ is approximatively the
radius of the nucleus and, in turn,
\begin{eqnarray}
\nu(T_e, Z) &=& \alpha_{\rm e.m.} Z E_e/p_e\nonumber\, ,\\ 
\gamma_0 &=& \sqrt{1-\alpha_{\rm e.m.}^2 Z^2}\nonumber\, ,
\end{eqnarray}
with $A$ and $Z$ atomic number and charge of the parent nucleus
respectively. The $Q$ values for the two $\beta$ sources are 1162.1~KeV
for the $^{210}$Bi and 687.1~KeV for the $^{85}$Kr. Moreover, since the
$\beta$ decay of the $^{85}$Kr is forbidden, the spectrum should be
corrected by the multiplicative factor $p_e^2/Q^2+(1-T_e/Q^2)^2$
\cite{Fermicorrected}. Of course the observed $\beta$ spectrum are the
convolution of the true spectrum with the resolution function.

\section{Analysis}\label{sec:analysis}

The observed total spectrum in Borexino is given by
\begin{equation}\label{eq:spectrum}
S(K_e)=\sum_s {\cal N}_s\Phi_s\int dE_\nu \varphi_s(E_\nu)
\sum_a\frac{d\tilde\sigma_a}{dT_e}(E_\nu,K_e)\cdot P_{ea}(E_\nu)
+\tilde{S}_\alpha(K_e)+\tilde{S}_\beta(K_e)\, ,
\end{equation}
where $s\in\{$pp, pep, $^7$Be, $^{16}$O, $^{14}$N$\}$ are the solar
sources (the contribution from $^8$B and hep neutrinos is negligible)
with their (normalized) spectra $\varphi_s$ and Standard Solar Model
flux $\Phi_s$ (in particular, we have used the AGS05 fluxes, see Table~6
of \cite{Bahcall:2005va}), $P_{ea}(E_\nu)$ is the oscillation
probability $P(\nu_e\to\nu_a)$, and ${\cal N}_s$ are normalization
factors (with the constraint ${\cal N}_s\geq 0$); the ``tilde'' means
that the cross sections/background spectra have been convoluted with the
resolution function. In practice the CNO sources (i.e., the $^{16}$O,
$^{14}$N) give a small contribution to the spectrum. Moreover, their
spectra are almost indistinguishable among them and from those of the
$^{210}$Bi background. Since we do not expect a strong spectral
distortion from oscillations, following the Borexino paper, we have
embedded the CNO contribution into the $^{210}$Bi spectrum.

For $^7$Be and pep neutrinos the spectrum is almost monoenergetic apart
a few keV broadening due mainly to collision and thermal effects
\cite{Bahcall:1994cf}. In the energy range explored by the experiment
the only contribution to the spectrum comes from the 863~keV $^7$Be
line, the contribution from the other line (385 KeV) being well below
the threshold. In our analysis we fix only the pp and pep neutrinos at
their Standard Solar Model value (i.e., we force ${\cal N}_{\rm
pp,pep}=1$) \cite{Bahcall:2005va}, as their sources are affected by a
very low theoretical uncertainty ($\sim$1-2\%). All the other solar
fluxes, the background normalizations and, of course, the non--standard
parameters, are taken as free variables in the fit.

Regarding the oscillation probability, we have used the standard
Mikheyev-Smirnov-Wolfenstein (MSW) probability with the oscillation
parameters defined in Eq.~(\ref{eq:massmixing}).%
\footnote{We have used the approximate formulae for calculating the
$P_{ee}$ survival probability already averaged over the production 
zone given in~\cite{deHolanda:2004fd}.}
We do not fit the mass-mixing parameters with Borexino data since it is
not the goal of this work. A comment is in order. In principle when non
standard interactions are present, also the oscillation probability
could be affected by new interactions. In particular, neutrino magnetic
moment can induce resonant spin-flip precession \cite{Pulido:1991fb} 
(for a recent review, see also \cite{Strumia:2006db}, Sec.~13.1 
and reference therein).
Anyway, barring oscillations into sterile neutrinos, any change in the
probability would reflect only in a very small change of the spectral
shape (since it is determined mainly by the $^7$Be line), while the
overall normalization is left free.%
\footnote{Moreover, to simplify our analysis, we do not allow any
$\nu\leftrightarrow\bar{\nu}$ transitions so that $\bar\nu$'s are absent
in the flux.}
For this  reason the precise choice of the probability is not critical
to our scope, provided that: 1) no active-sterile oscillations are
allowed and 2) no dramatic spectral distortion in the CNO spectra are
expected.

We now have all the ingredients to calculate the $\chi^2$:%
\footnote{In presence of bins with zero or few events the $\chi^2$
function should in principle be corrected as prescribed in \cite{PDG}.
However, since the low statistic bins are less relevant for the
analysis, we prefer to use the standard $\chi^2$. In this way the
unknown parameters can be extracted analytically.}
\begin{equation}\label{eq:chi2}
\chi^2 = \sum_{ij}(N^{\rm Th}_i-N^{\rm Exp}_i)(\sigma^{-2})_{ij} (N^{\rm
Th}_j-N^{\rm Exp}_j)\, ,
\end{equation}
where $N^{\rm Exp}_i$ ($N^{\rm Th}_i$) is the experimental (theoretical)
number of events in the $i$-th bin, and, as explained in the previous
section, the matrix $\sigma^2_{ij}$ is calculated as:
\begin{equation}
\sigma^2_{ij}=[N^{\rm Exp}_i+(0.15N^{\rm Exp}_i)^2]\delta_{ij}+
(0.25N^{\rm Exp}_i)\cdot (0.25N^{\rm Exp}_j) \, .
\end{equation}
Since all the unknown parameters appear linearly in the calculation of
$N^{\rm Th}_i$, the $\chi^2$ minimization is straightforward. Defining 
\begin{equation}
F^k_i=\frac{\partial N^{\rm Th}_i}{\partial {\cal N}_k}
\end{equation}
(with $k\in\{$pp, pep,$^7$Be, CNO+$^{210}$Bi, $^{85}$Kr,
$^{210}$Po$\}$), it is easy to show that
\begin{equation} 
{\cal N}_k=\sum_i\left[({\mathbf F}\sigma^{-2}{ \mathbf
F}^T)^{-1}{\mathbf F}\sigma^{-2}\right]_{ki} \left(N_i^{\rm
Exp}-\sum_{s\in\{{\rm pp, pep}\}}{\cal N}_s F^s_i\right)\, ,
\end{equation}
where the pp and pep sources are fixed and not fitted. In any case,
negative ${\cal N}_k$'s are not allowed.

\begin{figure}[t]
\centering \epsfig{file=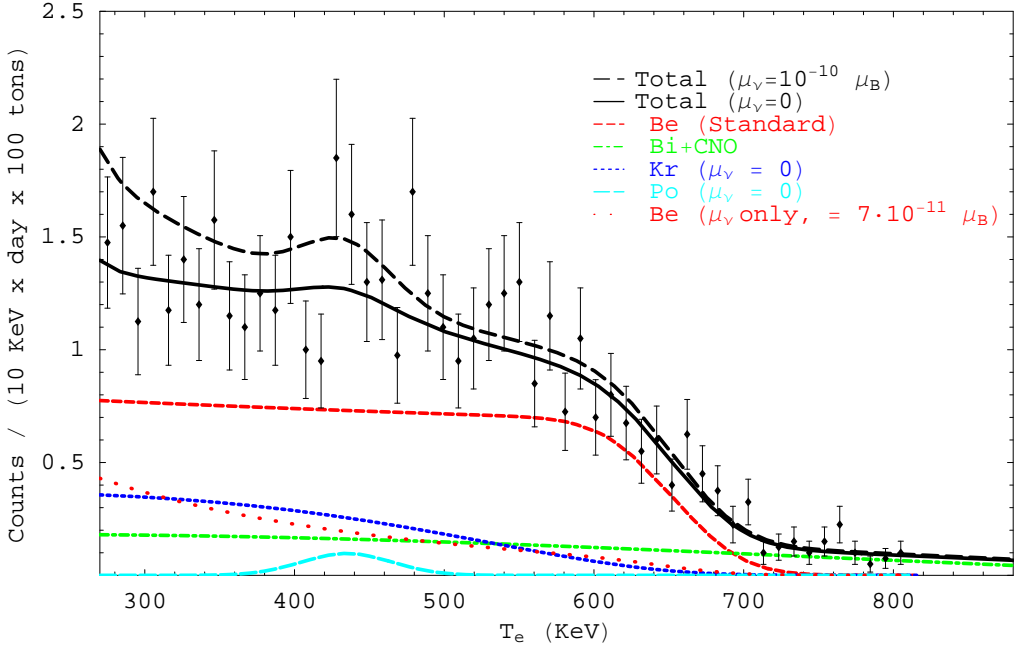, width=14truecm} 
\caption{(color online) The ``best fit'' spectrum with zero (black solid
line) and nonzero (black dashed line) magnetic moment. The contributions
of the $^7$Be source (red medium--dashed line) and those from the
CNO+$^{210}$Bi (green dotted--dashed line), $^{85}$Kr (blue dotted
line), and $^{210}$Po (light blue log-dashed line) in the case
$\mu_\nu=0$ are also shown. The number of counts per day in the ``best
fit'' ($\mu_\nu=0$) case in the whole energy range are 49 for the
$^7$Be, 12 for the CNO+$^{210}$Bi, 18 for the $^{85}$Kr, and 1 for the
$^{210}$Po source. For illustration, we also show the contribution coming 
from the magnetic moment only for $\mu_\nu=7\times 10^{-11}\mu_B$ (red dotted 
line).}
\label{fig:spectrum}
\end{figure}
The ``best fit'' spectrum with standard interactions only is shown in
Fig.~\ref{fig:spectrum}, which is similar to Fig.~6 of
\cite{Collaboration:2007xf}, with black solid line. (In the figure is
shown also the spectrum with a non zero magnetic moment with the black
dashed line). For comparison, we also report the number of events per
day in the full recoil energy range for each source for 100~Tons of
scintillator in our ``best fit'' case: 49 for the $^7$Be, 12 for the
CNO+$^{210}$Bi, 18 for the $^{85}$Kr, and 1 for the $^{210}$Po.

The value of $\chi^2_{\rm min}$ is 37.6, is slightly lower than the one
obtained by the collaboration ($\chi^2_{\rm min}=41.9$), due to
different assumptions in the two analyses. In the same figure the
various contributions (apart those from from pp and pep neutrinos which
are very small) are also shown. The main contributions to the spectrum
come from the CNO+$^{210}$Bi, $^{85}$Kr, and $^{210}$Po. Among these,
only the first could be slightly affected by the functional form of
$P_{ea}(E_\nu)$. We see also that the this contribution as almost flat,
while those from $^{210}$Po is a peculiar ``bump''. The main
contribution for the spectral distorsion at low energies (thus mimicking
those coming from non--standard interactions) comes from the $^{85}$Kr
background. As we will discuss later, this background is the main
limitation to the measure.

\section{Limits on the non standard interactions}\label{sec:results}

Introducing the non standard interactions, we see that the main
contribution to the spectral distortion comes from $^7$Be neutrinos. In
particular, we see that, if only $^7$Be neutrinos are taken into
account, the contribution to the spectrum coming from the non standard
interactions is:
\begin{equation}\label{eq:spectrum_mag}
\delta S(K_e)=\xi_{\rm eff}^2 \cdot \frac{d\tilde\sigma^{\rm non\,
std}}{dK_e}(E_\nu,T_e)\, ,
\end{equation}
where $\xi\equiv\mu_\nu\, , \lambda_\nu$ and $d\tilde\sigma^{\rm non,
std}/dK_e$ is the non standard contribution given by
Eqs.~(\ref{eq:mucontib}) or (\ref{eq:unpartcs}), after the proper
convolution with the resolution function. The label ``eff'' means that
we must consider an effective coupling, given by:
\begin{equation}\label{eq:effective}
\xi_{\rm eff}^2 = \sum_a P_{ea}(E^0_\nu) \cdot \xi_a^2 \, ,
\end{equation}
where $E^0_\nu=863$~keV is the energy of the $^7$Be neutrinos. Due to
the unitarity of the probability, this is also equivalent to introducing
an equal magnetic moment ($\mu_\nu$) or unparticle coupling
($\lambda_\nu$) to all flavors. Since also non-monochromatic sources
contribute to the spectra, for simplicity we stick into this simplified
hypothesis of equal non--standard parameters. However, since the final
result of the analysis does not depend critically on the functional form
of $P_{ea}$, the limits on $\mu_\nu$ and $\lambda_\nu$ are practically a
limit on $\mu_{\rm eff}$ and $\lambda_{\rm eff}$.

In particular, if we trust that the oscillation is simply given by the
MSW effect with small $\theta_{13}$, 
we have that $\mu_{\rm eff}^2$ and $\lambda_{\rm eff}^2$ are given by
\begin{equation}\label{eq:eff}
\mu_{\rm eff}^2 = P^{2\nu}(E^0_\nu) \cdot \mu_e^2
+\left[1-P^{2\nu}(E^0_\nu)\right] \left(\cos^2\theta_{23} \cdot
\mu_\mu^2+ \sin^2\theta_{23}\cdot \mu_\tau^2\right)
\end{equation}
(and, of course, a similar expression holds for $\lambda_{\rm eff}$). 
In principle, in the previous equation also interference terms 
among oscillation amplitudes can be present. However, with the 
oscillation parameter given in Eq.~(\ref{eq:massmixing}) these terms
average out and can be safely neglected \cite{Vogel:1989iv,Grimus:2002vb}.%
\footnote{Note that our definition of $\mu_a$ is different from those used 
in ref.~\cite{Grimus:2002vb}.}
In Eq.~(\ref{eq:eff}) $P^{2\nu}(E^0_\nu)$ is the two--neutrino
$P(\nu_e\to\nu_e)$ standard MSW probability and $\sin^2
\theta_{23}=0.38\div 0.63$ ($2\sigma$) from recent atmospheric and
accelerator (K2K and MINOS) analysis \cite{Schwetz:2007my}. Moreover,
for low energy neutrinos we have also that with good approximation
$P^{2\nu}\simeq \cos^2\theta_{12}\simeq 1/2$. 

\begin{figure}[t]
\centering \epsfig{file=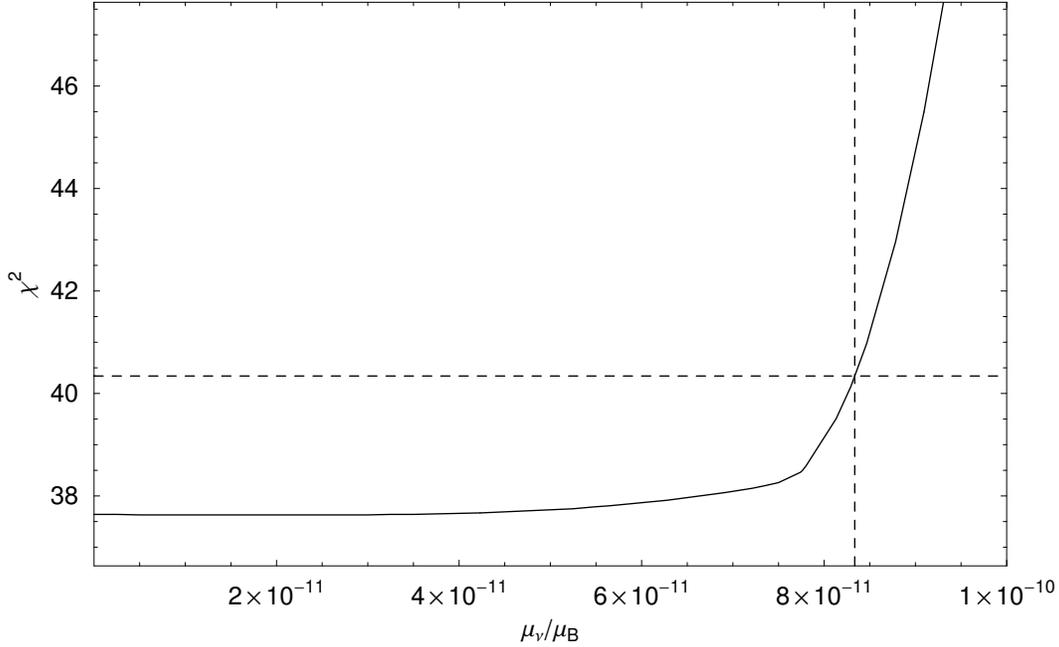,width=14truecm} 
\caption{The $\chi^2$ as a function of the neutrino magnetic moment
$\mu$ for standard axial and vectorial coupling. We also report the
$90\%$ C.L.\ ($\chi^2-\chi_{\rm min}^2 = 2.71$).} 
\label{fig:chi2_mag}
\end{figure}
\subsection{Limits on the magnetic moment}

We start by deriving the upper bound on the effective magnetic moment
from the SuperKamiokande (SK) experiment. To this end, the flux measured
by SK is:
\begin{equation}\label{eq:SKrate}
\Phi_{\rm SK} = \Phi_{^8{\rm B}} \frac {\int dE_\nu \varphi_{^8{\rm
B}}(E_\nu) \sum_a P_{ea}(E_\nu)\left[\sigma_a^{\rm
std}(E_\nu)+\mu^2_\nu\sigma^{\mu}(E_\nu)\right]} {\int dE_\nu
\varphi_{^8{\rm B}}(E_\nu) \sum_a P_{ea}(E_\nu)\sigma_e^{\rm
std}(E_\nu)} \, ,
\end{equation}
where $\sigma_a^{\rm std}$ and $\sigma^{\mu}$ are the total (standard
and non--standard) cross sections in the SK energy range. In this case
is not possible to define an effective magnetic moment as in
Eq.~(\ref{eq:eff}), so, for simplicity, we have assumed an equal
$\mu_\nu$ for all flavors.

From Eq.~(\ref{eq:SKrate}) we realize that the bound on $\mu_\nu$
depends on the total $^8$B flux (and hence is solar model dependent).
For example, using the AGS05 model \cite{Bahcall:2005va} where
$\Phi_{^8{\rm B}}=4.51\times 10^6\times (1\pm 0.12)$~cm$^{-2}$s$^{-1}$
with the SK measured rate $R_{\rm SK}=[2.35\pm 0.02$ (stat.)$ \pm 0.08$
(sys.)$]\times 10^6$ cm$^{-2}$s$^{-1}$ we obtain $\mu_\nu<2.1 \times
10^{-10}\mu_B$ (90\% CL). On the other hand for the GS98
\cite{Bahcall:2005va} model with the higher $^8$B flux $\Phi_{^8{\rm
B}}=5.69\times 10^6\times (1\pm 0.16)$~cm$^{-2}$s$^{-1}$ we obtain a
stronger bound, $\mu_\nu<1.3 \times 10^{-10}\mu_B$ (90\% CL), which is
more in agreement with the bound obtained by the spectral analysis done
by the SK collaboration \cite{Liu:2004ny}.

As far as Borexino is concerned, the corresponding upper bound on the
neutrino magnetic moment is not dependent on the Standard Solar Model
assumed. In fact the $^7$Be normalization is extracted from the
experimental data. In order to show the effect of the magnetic moment,
in Fig.~\ref{fig:spectrum} in black dashed line we plot also the
theoretical spectrum for $\mu_\nu=10^{-10}\mu_B$ (which is well beyond
the 90\% limit). From the plot we see that (as expected) the spectrum
grows at low energies. This behaviour, as we discuss later, can be also
mimicked by the $^{85}$Kr background.

In Fig.~\ref{fig:chi2_mag} we show the $\chi^2$ as function of the
neutrino magnetic moment where all the free fluxes and backgrounds have
been marginalized. We see that the 90\% C.L.\ limit on the magnetic
moment is $\mu_\nu\leq 8.4\times 10^{-11} \mu_B$.%
\footnote{In \cite{Grimus:2002vb} a perspective analysis of the Borexino
experiment was performed. Although they obtained a more stringent limit
on a combination of the Majorana magnetic transition moments, they
assumed a fixed background. As we will see later, the main source of
uncertainty comes from the lack of knowledge of the true number of
background events.}
We have also tried to fit the neutrino charge radius and more in general
the neutrino vector and axial couplings $g_V$ and $g_A$, as proposed in
\cite{Berezhiani:2001rt}. Unfortunately the limits obtained with
Borexino are far from being competitive from those obtained by other
experiments \cite{Barranco:2007ej}. However, we have verified that,
varying $g_V$ and $g_A$ inside the allowed region(s) in Fig.~2 of
\cite{Barranco:2007ej} our limit on $\mu_\nu$ does not vary appreciably.

Although this limit is less competitive than those obtained from reactor
experiments like in GEMMA ($\mu_e<5.8\times 10^{-11}\mu_B$ at 90\% C.L.,
\cite{Beda:2007hf}), we should bear in mind that reactor experiments are
short baseline. For this reason, the limits obtained in these
experiments are essentially bounds on the $\mu_e$ component. Instead,
the limits coming from solar neutrinos can be translated into limits for
$\mu_\mu$ and $\mu_\tau$. In particular, we have seen that for Borexino
the limit on $\mu_\nu$ is practically a limit on $\mu_{\rm eff}$. Using
Eq.~(\ref{eq:eff}) and $P^{2\nu}\simeq 1/2$ we obtain a conservative
limit on $\mu_\tau$ (obtained in the worst case $\mu_e=\mu_\mu=0$ and
$\sin^2\theta_{12}$, $\cos^2\theta_{23}$ taken at their $2\sigma$
minimum allowed values):
\begin{equation}
\mu_\tau\lesssim 1.9\times 10^{-10}\mu_B\, ,
\end{equation}
which is three order of magnitude stronger than those quoted by the
Particle Data Group ($\mu_\tau<3900\times 10^{-10}\mu_B$ \cite{PDG}).
Equivalently, we get the limit 
\begin{equation} \mu_\mu\lesssim 1.5
\times 10^{-10}\mu_B\,. 
\end{equation}

The ``plateau'' in the $\chi^2$ for $\mu_\nu\lesssim 0.85\times
10^{-10}\mu_B$ is due to the partial compensation between the $^{85}$Kr
background and the magnetic moment contribution (we have also checked
that the other normalization factors are almost insensitive to the value
of $\mu_\nu$). In fact both the magnetic moment and the $^{85}$Kr, have
the same effect in the spectrum, i.e., to increase the slope at low
energies. Increasing $\mu_\nu$ the slope is kept almost constant if one
simultaneously decreases the normalization of the $^{85}$Kr background.
For example, this can be also seen in Fig.~\ref{fig:spectrum} from 
the comparison between the red dotted line (which is the contribution 
of the  magnetic moment to the electron spectrum due to $^7$Be neutrinos, 
for $\mu_\nu=7\times 10^{-11}\mu_B$) and the blue dotted line (which is the
contribution due to the $^{85}$Kr in absence of magnetic moment).
At $\mu_\nu \sim 0.85\times 10^{-10}\mu_B$ the $^{85}$Kr normalization
vanishes and compensation is no longer possible. This can also be
noticed in the abrupt discontinuity of the derivative of the $\chi^2$.

The main limitation for the measure of $\mu_\nu$ thus comes, not from
the limited statistics, but from the imprecise knowledge of the
$^{85}$Kr background. If in the future this background will be reduced
by a purification campaign,%
\footnote{We notice that due to the long mean-life of the $^{85}$Kr, the
fate and impact of this background is different than that of $^{210}$Po.}
the limit on $\mu_\nu$ would become much stronger. We also mention that
by improving a lot the exposure it might be possible to measure the
$^{85}$Kr contamination by means of correlated events as reported in
\cite{Collaboration:2007xf}. Moreover, due to the Earth orbital
eccentricity, the solar flux has periodical variations in one year,
while the internal background is expected to be nearly constant in time.
This means that a better limit on $\mu_\nu$ could be obtained simply
measuring the spectrum in different periods of the year.

\begin{figure}[t]
\centering \epsfig{file=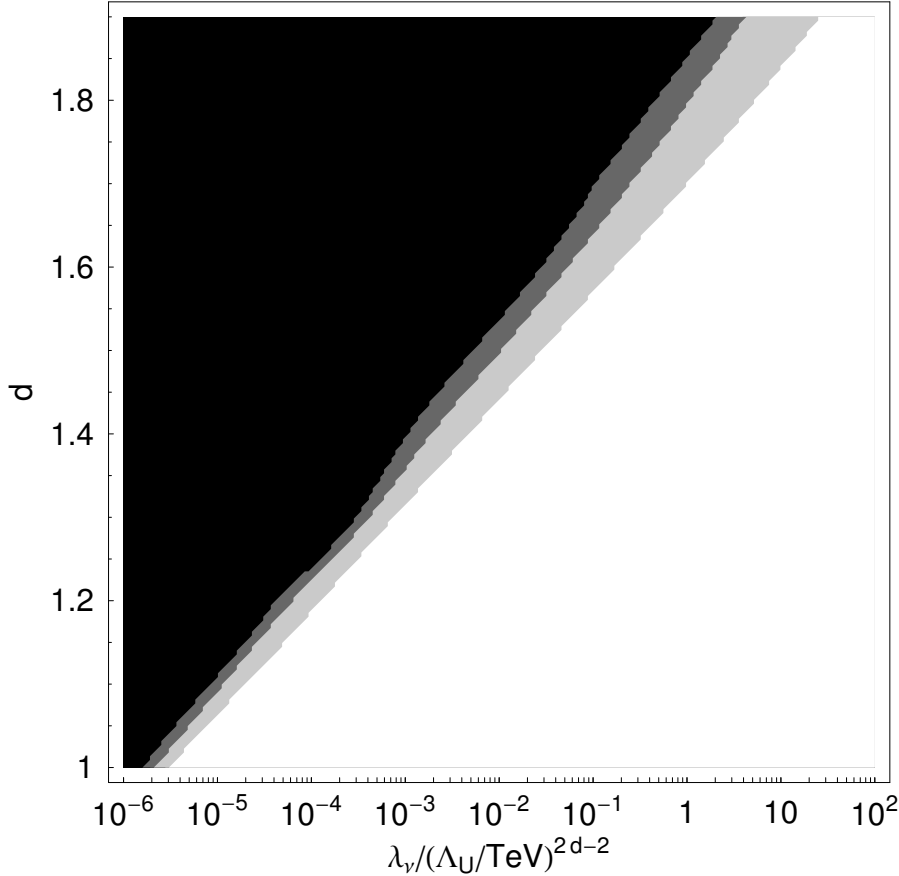,width=12truecm} 
\caption{The 90\% C.L.\ allowed zone in the plane $(d,\lambda_\nu)$ (for
$\Lambda_{\cal U}=1$~TeV) as from Ref.~\cite{Balantekin:2007eg} (light
gray), with unconstrained $^7$Be flux (dark gray) and with the $^7$Be
flux fixed by the Standard Solar Model. See the text for more details.}
\label{fig:unpart_allow}
\end{figure}

\subsection{Limits on unparticle coupling}

We have done the analysis done in the previous section but using the
unparticle cross section (\ref{eq:unpartcs}). All the considerations
made in the previous section (in particular, the partial compensation
between the $^{85}$Kr background and the unparticle contribution) apply
also in this case.  

In Table~\ref{tab:unparticles} we show our limits on $\lambda_\nu$ as
function of the dimension $d$. We fix the value of $\Lambda_{\cal U}$ to
1~TeV. Of course, if we choose a different value for $\Lambda_{\cal U}$,
the limits on $\lambda_\nu$ should be rescaled according to
Eq.~(\ref{eq:unpartcs}). The values of $d$ (first column) have been
chosen in order to make a comparison with the limit obtained in Table~1
of \cite{Balantekin:2007eg}. In the third column we show our 90\% C.L.\
limits on $\lambda_\nu$ obtained in the same way of the previous
section, i.e., taking the $^7$Be solar flux completely unconstrained. In
the fourth column we show the same limit, but taking into account the
10.5\% theoretical uncertainty on the $^7$Be flux \cite{Bahcall:2005va}.
This is done simply adding a penalty function to the $\chi^2$
\begin{equation}
\chi^2\rightarrow\chi^2+\left(\frac{{\cal N}_{^7{\rm Be}}-1}
{\sigma_{^7{\rm Be}}}\right)^2\, .
\end{equation}
As expected, in this case the bounds on $\lambda_\nu$ slightly improve.%
\footnote{We have checked that in the case of the magnetic moment the
improvement is negligible.}
In Fig.~\ref{fig:unpart_allow} we show also the 90\% C.L.\ allowed zone
in the plane $(d,\lambda_\nu)$ as from Ref.~\cite{Balantekin:2007eg}
(light gray), with unconstrained (dark gray) and constrained (black)
$^7$Be flux.

\begin{table}[t]
\centering
\begin{tabular}{ccccccc}
\hline\hline
$d$ & & 
$\lambda_\nu$ (Table~1 of \cite{Balantekin:2007eg}) & &
$\lambda_\nu$ (unconst.\ $^7$Be) & &
$\lambda_\nu$ (SSM $^7$Be)\\
\hline\hline
1.01 & & $3.5\times 10^{-6}$ & & $2.4\times 10^{-6}$ & & $1.8\times 10^{-6}$ \\
1.05 & & $7.3\times 10^{-6}$ & & $4.8\times 10^{-6}$ & & $3.4\times 10^{-6}$ \\
1.1  & & $1.9\times 10^{-5}$ & & $1.1\times 10^{-5}$ & & $0.8\times 10^{-5}$ \\
1.2  & & $1.2\times 10^{-4}$ & & $0.6\times 10^{-4}$ & & $0.4\times 10^{-4}$ \\
1.3  & & $7.2\times 10^{-4}$ & & $4  \times 10^{-4}$ & & $2.9\times 10^{-4}$ \\
1.4  & & $4.5\times 10^{-3}$ & & $1.8\times 10^{-3}$ & & $1  \times 10^{-3}$ \\
1.5  & & $2.7\times 10^{-2}$ & & $1  \times 10^{-2}$ & & $0.5\times 10^{-2}$ \\
1.7  & & $9.5\times 10^{-1}$ & & $2.4\times 10^{-1}$ & & $1  \times 10^{-1}$ \\
1.9  & & $24.5             $ & & $4                $ & & $2$                 \\
\hline
\end{tabular}
\caption{Confrontation between our 90\% C.L.\ limits on $\lambda_\nu$
with $^7$Be unconstrained (third column) and fixed by the SSM fourth
column for different value of $d$, with  $\Lambda_{\cal U}=1$~TeV. Also
the limits obtained in \cite{Balantekin:2007eg} are shown (second
column).}\label{tab:unparticles}
\end{table}

We see that our limits are even stronger than those obtained by
Balantekin and Ozansoy \cite{Balantekin:2007eg}. We remark that the
limits obtained in \cite{Balantekin:2007eg} are obtained using short
baseline reactor data thus sensitive only to $\nu_e$, while our result
applies to the combination $\lambda_{\rm eff}$ in
Eq.~(\ref{eq:effective}). Since $P_{ee}\sim 1/2$, sometimes our limits
on $\lambda_e$ alone can be weaker than those in
\cite{Balantekin:2007eg}. However, combining our limits on $\lambda_{\rm
eff}$ with those on $\lambda_e$ in \cite{Balantekin:2007eg} we can
obtain for the first time bounds on the coupling constants $\lambda_\mu$
and $\lambda_\tau$ according with Eq.~(\ref{eq:eff}).

A comment is in order. In principle, unparticle interactions could affect 
also the production and propagation of neutrinos \cite{GonzalezGarcia:2008wk}.
However, we have already stressed that the major contribution of the solar
flux comes  from the $^7$Be monoenergetic line. This largely compensate eventual 
uncertainties in the conversion probability. For this reason, we think
that the analysis with the completely unconstrained $^7$Be 
should be considered to be more reliable.

Recently, in \cite{Grinstein:2008qk} it has also pointed out that a
generic unparticle scenario will generate contact interactions between
the particle operators. This contact term would generate a scattering
amplitude similar to those in Eq.~(\ref{eq:unpscat}), but with $d=2$ and
${\cal F}^2=1$.%
\footnote{We thank B.\ Grinstein for pointing us this fact.}
Assuming that this term is dominant (and thus any interference term is
negligible), the $\nu$-$e$ cross section is thus similar to those in
Eq.~(\ref{eq:unpartcs}) but with 
\begin{equation}
\lambda^{\rm cont}_a=\left[\sum_b (\lambda_{e\nu}^{a b})^2\right]^{1/2}\, ,
\end{equation}
where $\lambda_{e\nu}^{a b}$ are the contact term couplings for the
$\nu_a e\to\nu_b e$ scattering. The equivalent limit on $\lambda^{\rm
cont}_\nu$ for $\Lambda_{\cal U}=1$~TeV is $\lambda^{\rm cont}\leq 3.5$
($1.4$) in the hypothesis that $^7$Be neutrino flux is unconstrained
(constrained). We do not perform a combined analysis with $\lambda_\nu$
and $\lambda^{\rm cont}_\nu$ unconstrained. However, since we see that
the limit on is $\lambda^{\rm cont}_\nu$ in generally much weaker than
those obtained for $d<2$, if the two coupling constants were of the same
magnitude, the contribution to the cross section due to the non--contact
term would be dominant. In this hypothesis, the limits on $\lambda_\nu$
in Fig.~\ref{fig:unpart_allow} can be safely assumed.

\section{Conclusions}\label{sec:conclusions}

In this paper we have analyzed the first Borexino data release to
constrain the neutrino magnetic moment and the coupling of neutrino and
electrons with an hypothetical unparticle sector. The analysis is
performed analyzing the spectrum of the recoil electron energy. Since
the leading contribution to this spectrum comes from the monoenergetic
solar $^7$Be neutrinos, the shape of the spectrum is almost independent
from the energy dependence of the oscillation probability. The other
contribution to the spectral shape is due to the internal background of
the detector.

We have performed a $\chi^2$ fit assuming unconstrained both the solar
fluxes (except for the pp and pep, which, however, give a negligible
contribution) and the internal backgrounds. In absence of non--standard
effects our results are in good agreement with those obtained by the
Borexino collaboration.

An upper limit on the neutrino magnetic moment is found: $\mu_\nu\leq
0.85\times 10^{-10} \mu_B$. Although this is not the strongest limit in
the literature, we stress that the magnetic moment measured by Borexino
is a linear combination of the magnetic moments of the different
neutrino flavors. Within a reasonable assumption on the oscillation
probability this limit translates into an upper one for the (much more
unconstrained) $\mu_\tau$.

In the same way, we have obtained limits on the couplings with the
unparticle sector. The results are summarized in
Table~\ref{tab:unparticles}. In this case the limits are stronger than
those obtained in the literature \cite{Balantekin:2007eg}. Furthermore,
in this case the coupling constant is actually a linear combination of
the couplings of neutrinos with different flavors. In this way we have
obtained upper bounds on the still unknown parameters $\lambda_\mu$ and
$\lambda_\tau$. We want here to stress the fact that this limit strictly
applies if {\em only} scalar and contact unparticle operators are present. 
If other operators are included in the analysis the bounds would
be relaxed. A more complete analysis with further unparticle operators 
is deserved for a time being.

Finally, we are confident that in future Borexino will improve these
limits. Since the effect of non--standard interactions is partly hidden
by the internal contaminants, a reduction or at least a better knowledge
of these backgrounds would be of a great help. With at least one year of
data-taking, using the seasonal variations of the solar neutrino flux
due to the orbital eccentricity (about 3\% in one year) Borexino will be
able to disentangle the contribution to the spectrum coming from the
$^7$Be neutrinos to that due to the (if constant) background. Moreover,
a substantial reduction of systematics and a dramatic increase in the
statistics is also expected.

\section{Acknowledgements}

We are grateful to A.\ Ianni for fruitful discussions and suggestions
and for the careful reading of the manuscript. We also thank G.\
Ranucci, T.\ Schwetz, J.\ W.\ F.\ Valle and B.\ Grinstein 
for discussions. One of us (M.\ P.) would like to thank the organizers 
of PASC in Sesimbra where part of this work has been done. The work of 
D.\ M.\ is supported in part by the Italian ``Istituto Nazionale di Fisica 
Nucleare'' (INFN) and by the ``Ministero dell'Istruzione, Universit\`a e 
Ricerca'' (MIUR) through the "Astroparticle Physics" research project. 
J.\ P.\ is grateful to the University of Lecce for hospitality.


\end{document}